\documentclass[a4paper,11pt]{article}
\usepackage{jinstpub} 

\usepackage{lineno}

\usepackage{booktabs}
\usepackage{acronym} 

\acrodef{CMS}{Compact Muon Solenoid}
\acrodef{ECAL}{Electromagnetic Calorimeter}
\acrodef{EB}[EB]{ECAL barrel}
\acrodef{APD}{avalanche photodiode}
\acrodef{HL-LHC}[HL-LHC]{High-Luminosity LHC}
\acrodef{VFE}[VFE]{Very-Front-End}
\acrodef{CATIA}{CAlorimeter Trans-Impedance Amplifier}
\acrodef{LiTE-DTU}[LiTE-DTU]{Lisbon-Turino ECAL Data Transmission Unit}
\acrodef{ASIC}{application-specific integrated circuit}
\acrodef{ADC}{analogue-to-digital converter}
\acrodef{CERN}{European Organization for Nuclear Research}
\acrodef{ESS}{Environmental Stress Screening}
\acrodef{AFG}{Arbitrary Waveform Generator}
\acrodef{FPGA}{field-programmable gate array}
\acrodef{PLL}{Phase-Locked Loop}
\acrodef{DAC}{digital-to-analogue converter}
\acrodef{RMS}{root mean square}
\acrodef{TID}{total ionising dose}
\acrodef{LVR}{Low-Voltage Regulator}

\usepackage{siunitx} 
\DeclareSIUnit{\sample}{Sa} 

\title{CMS ECAL VFE Production Qualification}

\author[a,1]{C.~Borca,\note{Corresponding author.}}
\author[a,b]{S.~Argiro,}
\author[a, b,c]{L.~T.~Bulaja,}
\author[c,d]{D.~Brunet,}
\author[e]{F.~Cavallari,}
\author[a,b]{D.~Cristoforetti,}
\author[f]{G.~Dissertori,}
\author[a]{L.~Finco,}
\author[g]{M.~Kovac,}
\author[f]{B.~Kaynak,}
\author[f]{W.~Lustermann,}
\author[c]{P.~Milenovic,}
\author[a]{N.~Pastrone,}
\author[h]{P.~Rebello Teles}

\affiliation[a]{INFN Sezione di Torino,\\Via P.Giuria 1, 10125, Torino, Italy}
\affiliation[b]{Physics department, University of Torino,\\Via P.Giuria 1, 10125, Torino, Italy}
\affiliation[c]{Faculty of Physics, University of Belgrade,\\Pregrevica 118, 11080 Belgrade, Serbia}
\affiliation[d]{Faculty of Sciences, University of Novi Sad,\\Trg Dositeja Obradovića 3
21102, Novi Sad, Serbia}
\affiliation[e]{Sapienza University and INFN Roma I,\\Piazzale Aldo Moro 5, 00185, Roma, Italy}
\affiliation[f]{Institute for Particle Physics and Astrophysics, ETH Zurich,\\Otto-Stern-Weg 5, CH-8093 Zurich, Switzerland}
\affiliation[g]{University of Split, Faculty of Science,\\Ruđera Boškovića 33,
21000, Split, Croatia}
\affiliation[h]{Brazilian Center for Physics Research (CBPF), \\Dr.\ Xavier Sigaud, 150 Urca, 22290-180, Rio de Janeiro, RJ, Brazil.}

\emailAdd{cecilia.borca@cern.ch}

\abstract{
Upgrading the CMS ECAL barrel for the HL-LHC requires replacement of all on-detector electronics, including 12240 Very-Front-End (VFE) cards employing CATIA and LiTE-DTU ASICs for amplification and digitisation of signals from ECAL’s 61200 lead-tungstate crystals. 
All production VFEs undergo the same qualification procedure: functional testing, environmental stress screening, and calibration. We have received and functionally tested 13596 unique VFEs; two never passed our test criteria. Results for other VFEs are uniform and within expectations, demonstrating the excellent quality of the cards. Key metrics, such as a gain ratio of $10.21\pm0.07$ and power consumption of \qty{1.805 \pm 0.013}{\watt}, demonstrate high consistency across the batch.
}


\begin{document}
\notoctrue
\maketitle
\acresetall 
\flushbottom


\section{Introduction} 
\label{sec:intro} 

The \ac{CMS}~\cite{CMS_TP}\cite{CMS_Run3}
\ac{ECAL} barrel (\acs{EB})\acused{EB}
is a homogeneous calorimeter comprising
61200 lead-tungstate crystals, read out
by \acp{APD}~\cite{ECAL_TDR}.
For operation at the \ac{HL-LHC},
the crystals and
photodetectors will be retained, while
the complete on- and off-detector
readout electronics will be
replaced~\cite{CMS_Phase2_TDR}.
The upgrade is designed to preserve
\ac{EB}'s excellent energy resolution while
adding \qty{30}{\pico\second} timing
resolution for electrons and photons
with energies above
\qty{50}{\giga\electronvolt}, and to
increase the sampling frequency from
\qtyrange{40}{160}{\mega\sample\per\second}.

\begin{figure}[htbp]
    \centering
    \includegraphics[width=0.92\linewidth]{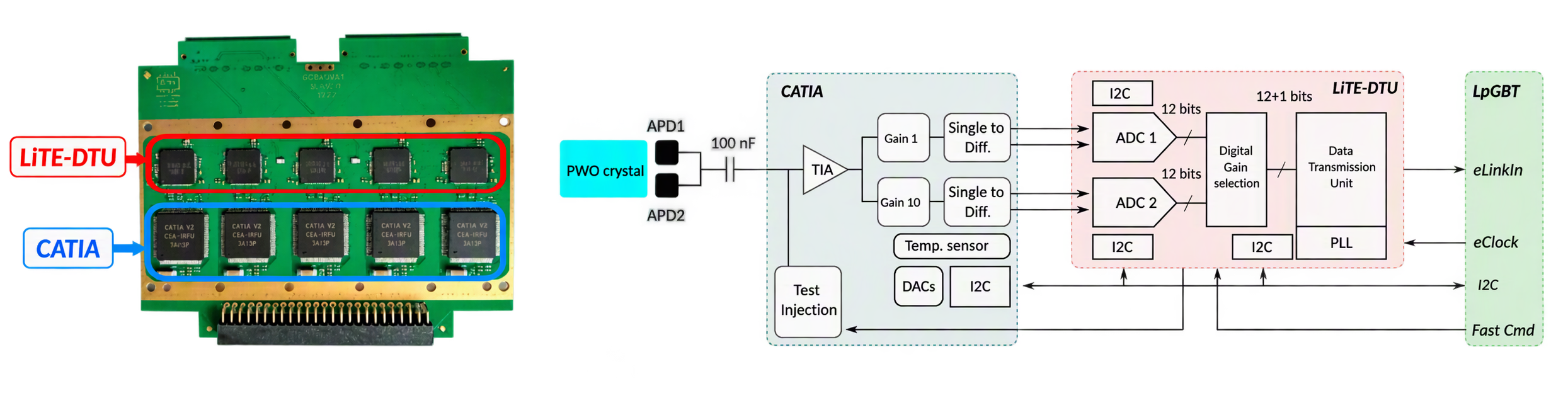}
    \caption{Photograph of the \acs{VFE} card (left), showing its five
readout channels, each equipped with a \acs{CATIA} and a
\acs{LiTE-DTU} \acs{ASIC}, and schematic representation
of a single readout channel (right), illustrating the
two parallel gain paths following the common
transimpedance stage.}
    \label{fig:vfe_and_schema}
\end{figure}

The \ac{VFE} card~\cite{VFE_design} hosts five
independent readout channels, each
consisting of a \ac{CATIA}~\cite{CATIA} and a
\ac{LiTE-DTU} \ac{ASIC}. \ac{CATIA}
provides two parallel
gain paths, with nominal gains of 1
and 10, allowing signals corresponding,
respectively, to energies up to
approximately \qty{2}{\tera\electronvolt}
and \qty{200}{\giga\electronvolt} to be
measured. The two signals are digitised
by two 12-bit \acp{ADC} at
\qty{160}{\mega\sample\per\second} in the
\ac{LiTE-DTU}, which also performs gain
selection, data compression and
serialisation at
\qty{1.28}{\giga\bit\per\second}%
~\cite{LiTEDTU}.
A picture of the \ac{VFE} card and a
schematic representation of each
readout channel can be seen in
Figure~\ref{fig:vfe_and_schema}.

This paper describes the \ac{VFE} production
qualification strategy and presents
the \acs{CERN} reception test results.
Of the 14200 production cards, 12240
will be installed in the \ac{EB}.


\section{Materials and Methods}
\label{sec:qualification} 

All production \ac{VFE} cards undergo a
multi-stage qualification process.
A first functional test, with a
slightly reduced scope, lasting
approximately \qty{30}{\second}, is
performed by the manufacturer,
allowing for a close
follow-up of the assembly process.
A second, complete functional test is
performed at reception at \ac{CERN}.
All cards then undergo \ac{ESS}~\cite{VFE_ESS},
followed by a final functional test
and gain calibration with an external
\ac{AFG}.

A single test system is used for all
qualification stages. One copy is
deployed at the manufacturer and two
at \ac{CERN}. The setup comprises
custom mechanics featuring fast
installation and removal of the
\acp{VFE}, dedicated adapter boards, a
\ac{FPGA}-based interface card
(VICE++)~\cite{VICE}, a four-channel power supply
and, where required, an \ac{AFG}.
A modular software suite selects the
test sequence for each qualification
stage; the only hardware difference
is the presence or absence of the
\ac{AFG}. Results and configuration
parameters are associated with the
unique \ac{VFE} identifier and stored
in a database, with larger datasets
kept for offline analysis.

At reception, each \ac{VFE} undergoes
an automated qualification sequence:
\begin{itemize}
\setlength{\itemsep}{0pt}
\setlength{\parsep}{0pt}
\setlength{\topsep}{0.5ex}
\item \textbf{Power consumption:}
  voltages and currents of both
  \ac{VFE} power rails are measured
  and the power consumption is
  computed.
\item \textbf{Temperature:}
  \ac{CATIA}  internal temperature sensors are read, as well as the FPGA temperature.
\item \textbf{\acs{PLL} scan / auto-lock:}
  the \ac{LiTE-DTU} autolock status
  bit is read via both I2C and 
  the dedicated \texttt{PLLLocked} pin.
  All \(7 \times 8\) combinations of
  the two \ac{PLL} control registers
  are then set, and the lock status is
  recorded at each setting; the
  locking range is deduced from this
  scan.
\item \textbf{$V_{\mathbf{ref}}$ scan:}
  \ac{CATIA} provides a 4-bit
  register to fine-tune the reference
  voltage $V_{\mathrm{ref}}$. All settings 
  are scanned, and the code closest to 
  $V_{\mathrm{ref}}=\qty{1}{\volt}$ is 
  selected from the corresponding auxiliary-ADC 
  reading, scaled by 0.8 to fit its dynamic range.
\item \textbf{\acs{ADC} calibration:}
  $V_{\mathrm{ref}}$ is sent from
  \ac{CATIA} to the \ac{LiTE-DTU},
  which is then triggered to perform
  the \ac{ADC} calibration internally.
\item \textbf{Pedestal optimisation:}
  the pedestal is the mean (DC) value
  of the \ac{CATIA} output. It is
  adjusted with a 6-bit \ac{DAC} in
  \ac{CATIA}. All settings are measured
  and the value closest to, and not
  below, 20 \ac{ADC} counts is
  selected. An 8-bit
  baseline-subtraction register in the
  \ac{LiTE-DTU} is then used to reach
  the value closest to 20 \ac{ADC}
  counts. The pedestal \ac{RMS} is
  the channel noise, evaluated for
  both gain paths.
\item \textbf{Internal test pulse:}
  \ac{CATIA} features a 12-bit
  test-pulse generator. A pulse of
  fixed amplitude is injected, the
  amplitude is reconstructed and
  checked to lie in the expected
  range.
\item \textbf{Internal linearity:}
  measurement of internal test pulses
  with 10 different amplitudes for
  low and high gain.
\item \textbf{Internal calibration:}
  the gain is measured with 10
  amplitudes of the internal
  test-pulse generator, covering the
  full high-gain range and the
  overlapping region of the low-gain
  path. For each amplitude and gain,
  40 points are recorded. Gains are
  obtained from linear fits to these
  data and used to compute the gain
  ratios. The measurement is performed
  for
  $R_{\mathrm{TIA}}=\qtylist{340;400;500}{\ohm}$
  with capacitor $C_1$ enabled,
  providing a \qty{35}{\mega\hertz}
  low-pass filter.
\end{itemize}
External calibration is performed
only after \ac{ESS} and is not
included in the results presented
here.


\section{Results}
\label{sec:results}

As of the end of October 2026,
13596 unique production \ac{VFE}
cards had been tested at \ac{CERN}
reception. 
The outcomes are summarised
in Table~\ref{tab:yield}.
Cards that passed all tests without a
recorded failure are distinguished
from those that failed at least one
test but later passed on retesting.
Of the 118 cards with at least one recorded 
failure, 116 later passed on retesting (0.85\%). 
DAC-scan anomalies accounted for approximately 
80\% of recorded test failures, while less 
frequent failures involved pedestal, linearity, 
and power-on checks. Only two cards never passed 
(0.01\%), both due to power anomalies.

\vspace{-3pt}
\begin{table}[htbp]
    \centering
    \vspace{-7pt}
    \caption{Reception test outcome for
    13596 unique production
    \acs{VFE} cards.}
    \label{tab:yield}
    
    \vspace{4pt}
    \small
    \begin{tabular}{|l|rr|}
        \hline
        \textbf{Category} &
        \textbf{Count} &
        \textbf{[\%]} \\
        \hline
        Valid cards &
        13596 & 100.00 \\
        Pass without recorded failure &
        13478 & 99.12 \\
        Pass with recorded failure &
        116 & 0.87 \\
        Never passed &
        2 & 0.01 \\
        \hline
    \end{tabular}
\end{table}

\begin{figure}[htbp]
    \centering
    \includegraphics[width=0.45\linewidth]{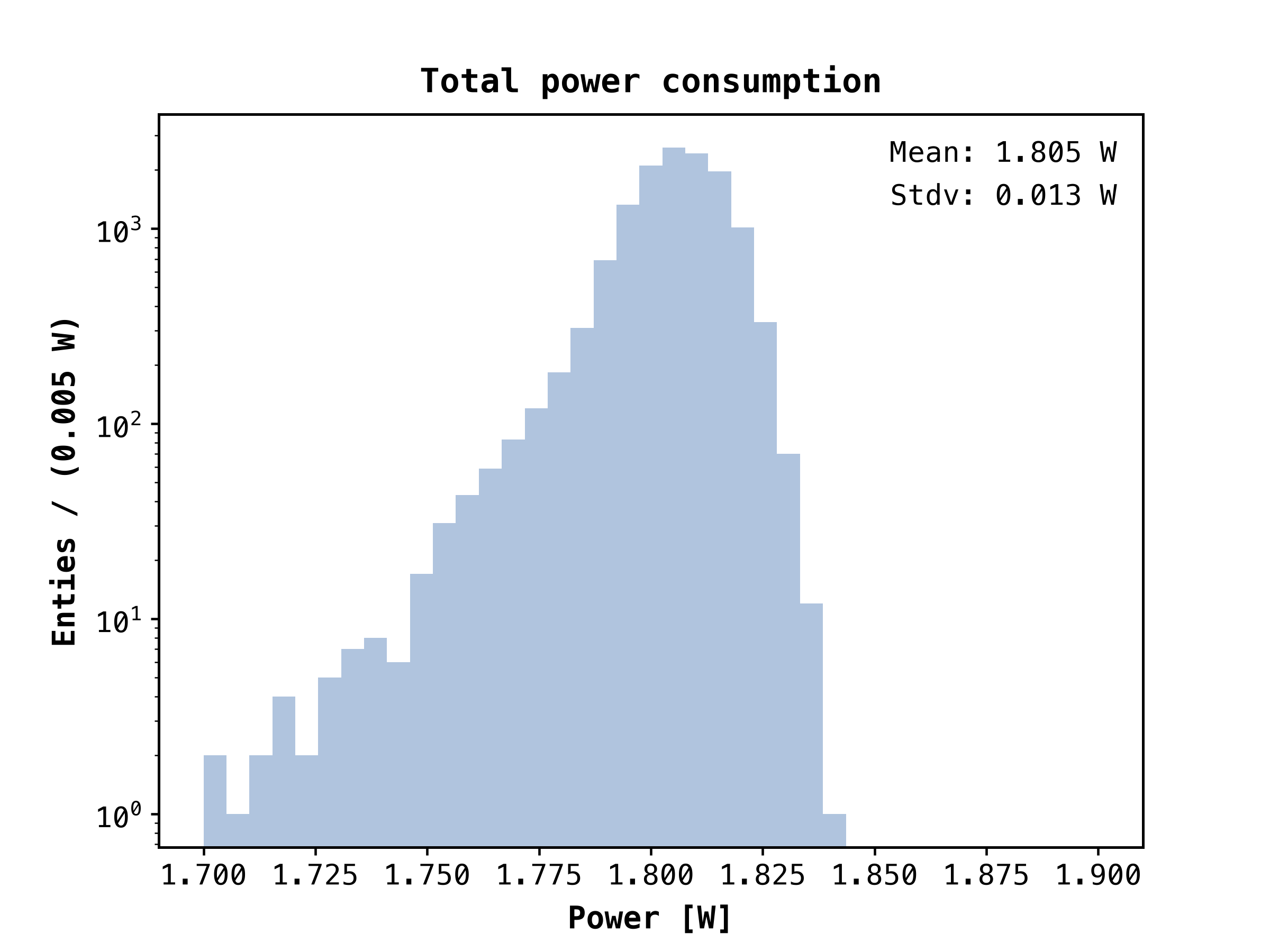}\quad
    \includegraphics[width=0.45\linewidth]{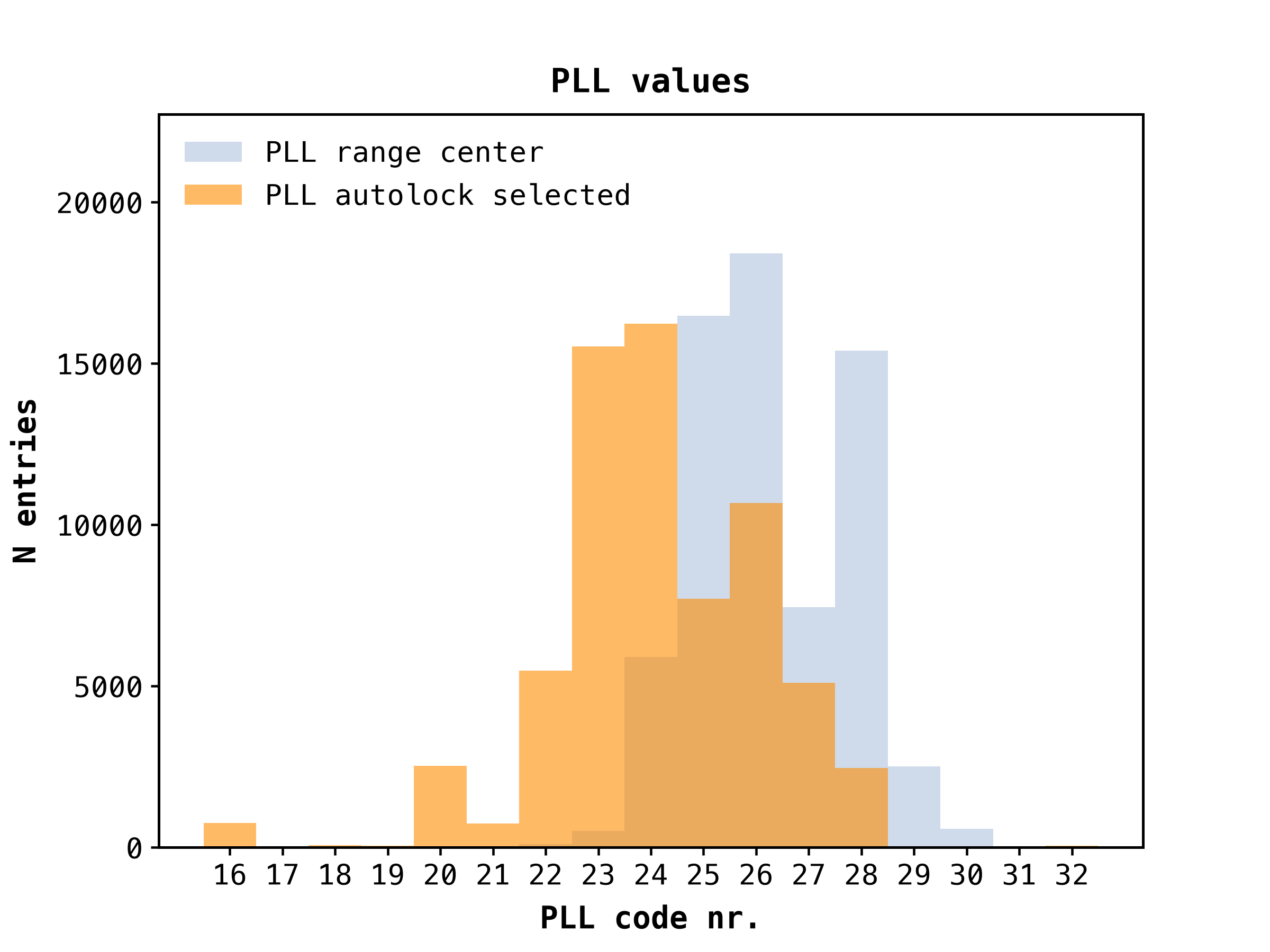}
    \caption{
    Distribution of the total \acs{VFE} power consumption
(left), shown on a logarithmic count scale, and
distributions of the \acs{PLL} scan centre and
autolock codes (right).}
    \label{fig:pll}
\end{figure}

The total \ac{VFE} power consumption
and the \ac{LiTE-DTU} \ac{PLL}
configuration are shown in
Figure~\ref{fig:pll}.
The power distribution has a mean of
\qty{1.805 \pm 0.013}{\watt}.
A logarithmic count scale highlights a 
low-power tail. 
This is not indicative of 
anomalous behaviour, as a similar feature 
has been observed and characterized in 
the \ac{PLL} RF power domain, which dominates
the LiTE-DTU power consumption.
The automatically selected lock code is 
expected to lie slightly ahead of the 
geometrical centre of the locking region. 
This behaviour, including its relation to the 
improved jitter performance in this region, 
was observed, studied, and discussed in 
detail in~\cite{LITEDTU_ARGIRO}. 
Consequently, the two distributions
exhibit a systematic shift.

The \ac{CATIA} internal temperatures
were recorded during each test.
Correct readings confirm the proper
functioning 
of the I2C interface and
the switch that routes
either the temperature sensor or
$V_{\mathrm{ref}}$ to the
measurement pins. 

\begin{figure}[htbp]
    \centering
    \includegraphics[width=0.45\linewidth]{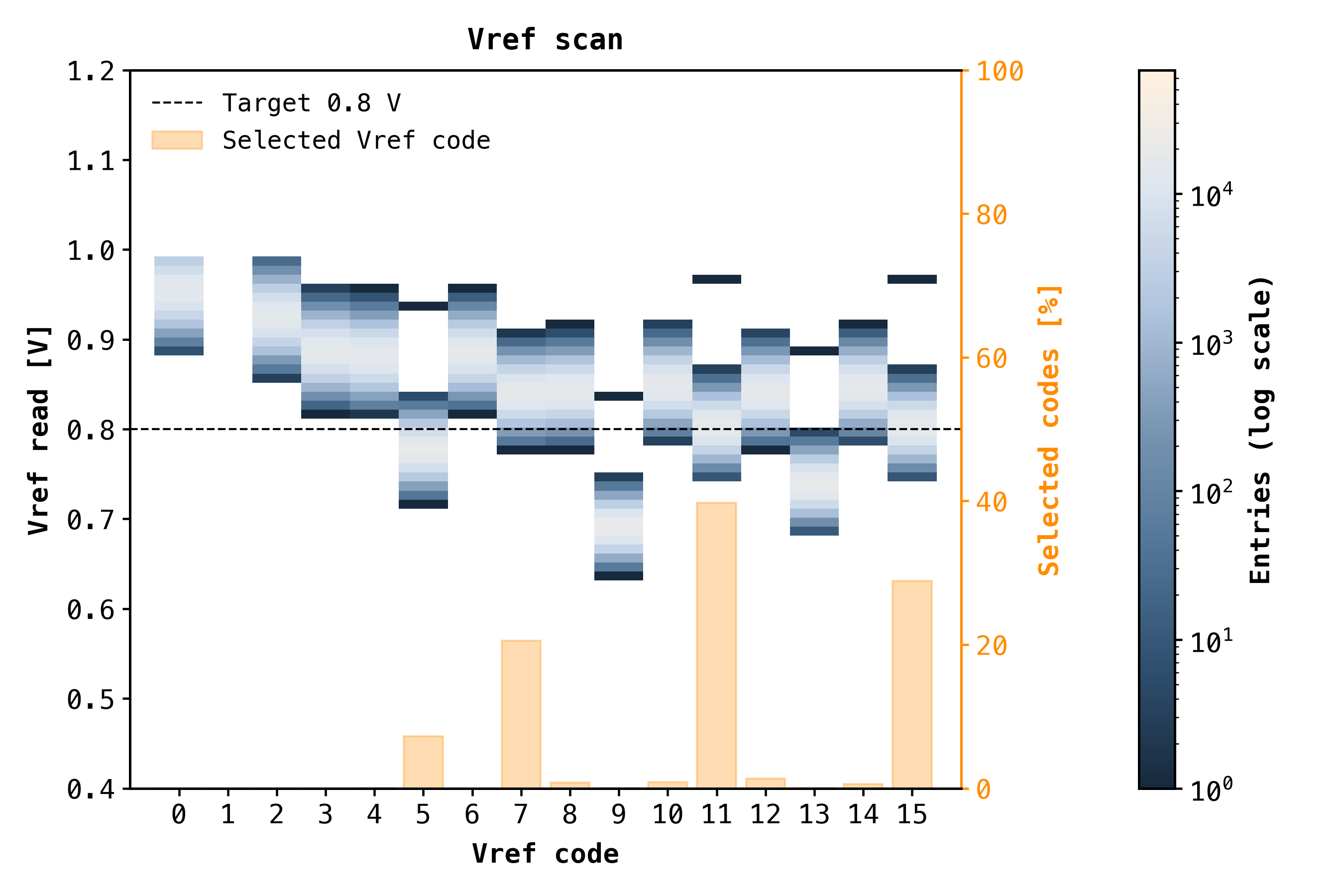}\quad
    \includegraphics[width=0.40\linewidth]{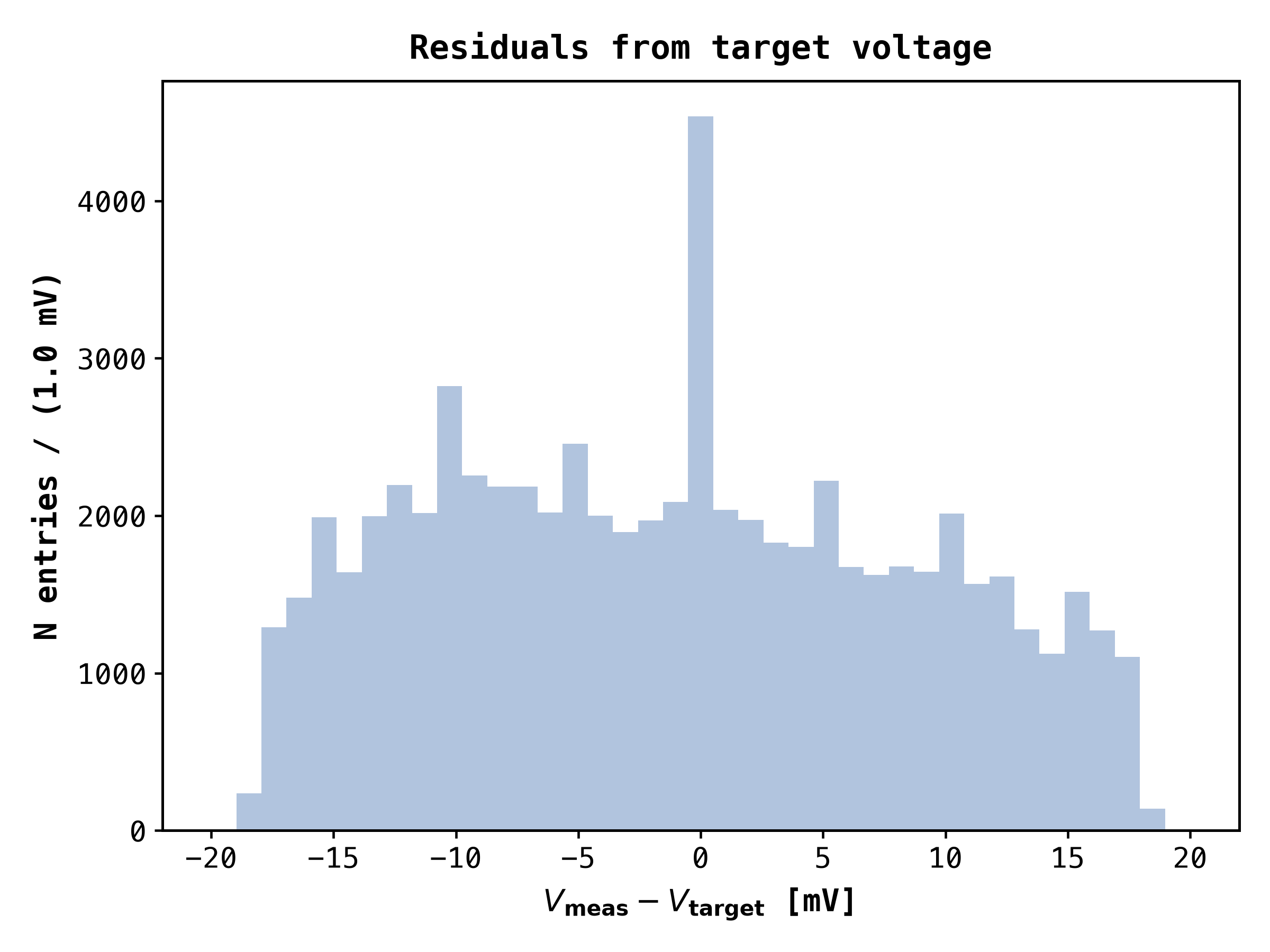}
    \caption{\ac{CATIA} reference voltage as a function of register
code (left), showing the \qty{0.8}{\volt} target and
the distribution of selected codes, and distribution
of the residuals relative to the target voltage (right).}
    \label{fig:vref}
\end{figure}

The \ac{CATIA} $V_{\mathrm{ref}}$ scan
is shown in Figure~\ref{fig:vref}.
The measured voltage as a function of
the 4-bit register code follows the
expected steps towards the
\qty{1}{\volt} (scaled by 0.8) target.
The distribution of the corresponding 
selected voltages has a spread of \qty{\pm20}{\milli\volt}, which corresponds to a maximum deviation of about 2.5\% from the target value. 

The \ac{LiTE-DTU} \acp{ADC} are
calibrated internally after the
identified $V_{\mathrm{ref}}$
settings have been programmed in
\ac{CATIA}.
The calibration succeeded for all
cards in the analysed sample.

\begin{figure}[htbp]
    \centering
    \includegraphics[width=0.95\linewidth]{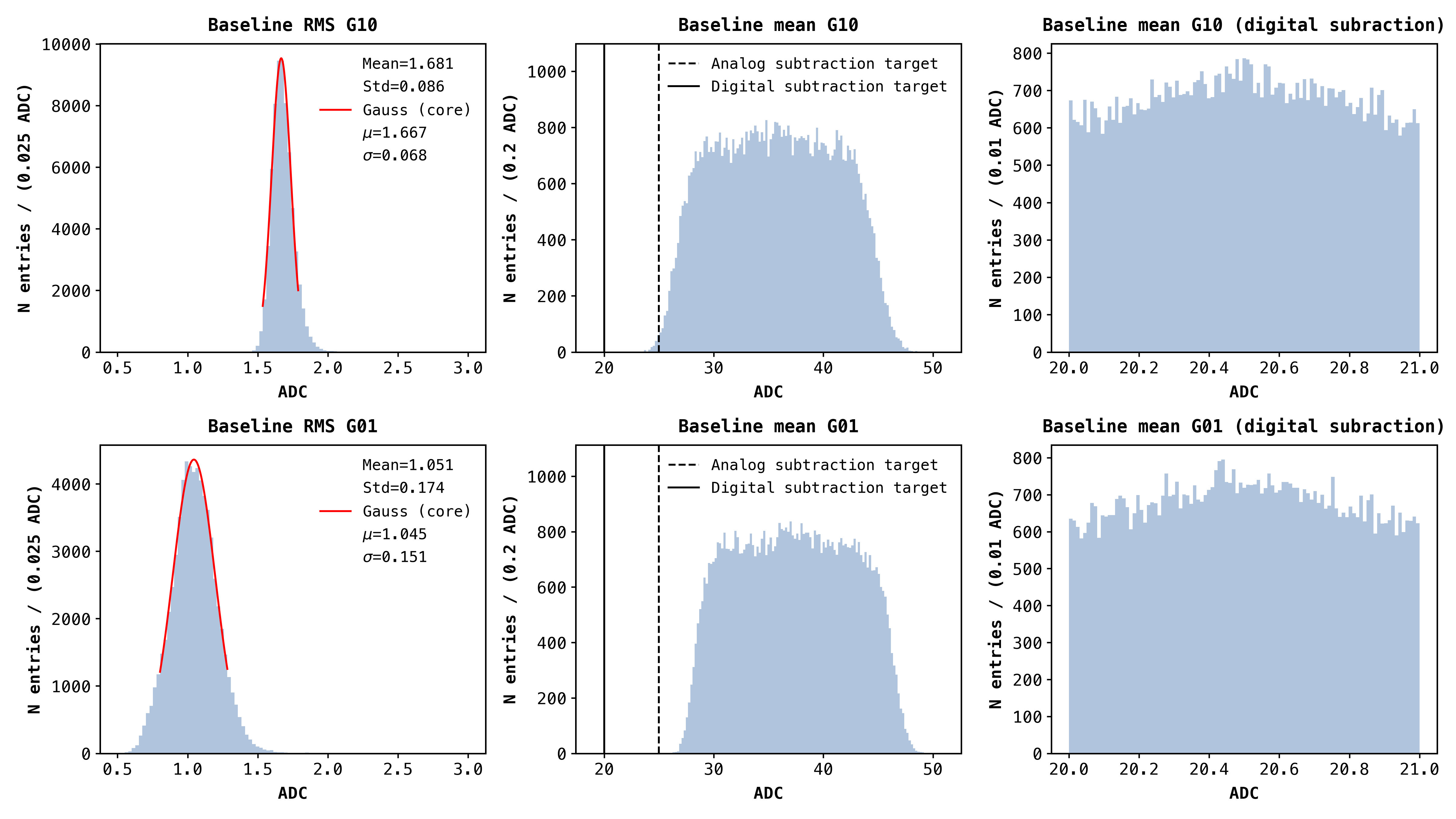}
    \vspace{-6pt}
    \caption{ Pedestal distributions for gain-10 (top row) and gain-1 (bottom row). From left to right: pedestal
\acs{RMS} with Gaussian-core fit, analogue pedestal following \acs{CATIA} \acs{DAC} optimisation, and residual pedestal after \acs{LiTE-DTU} digital subtraction.}   
    \label{fig:baselines}
\end{figure}

The two-stage pedestal optimisation is
shown in Figure~\ref{fig:baselines}.
After the \ac{CATIA} \ac{DAC}
selection the analogue pedestal still
spans several \ac{ADC} counts
(centre panels);
digital subtraction in the
\ac{LiTE-DTU} brings the baseline
close to 20 \ac{ADC} counts
(right panels).
The pedestal \ac{RMS} values
(channel noise) are
$\num{1.05 \pm 0.17}$ and
$\num{1.68 \pm 0.09}$ \ac{ADC}
counts for gain 1 and gain 10,
respectively (left panels).
Anomalies in the \ac{DAC} scan are the
most frequent recorded failure; they
typically occur far from the selected
operating point and do not imply
degraded performance there.

A fixed-amplitude pulse from the
\ac{CATIA} 12-bit test-pulse generator
is injected and decoded.
The reconstructed amplitude lay in the
expected range for all cards in the
analysed sample.

\begin{figure}[htbp]
    \centering
    \includegraphics[width=0.45\linewidth]{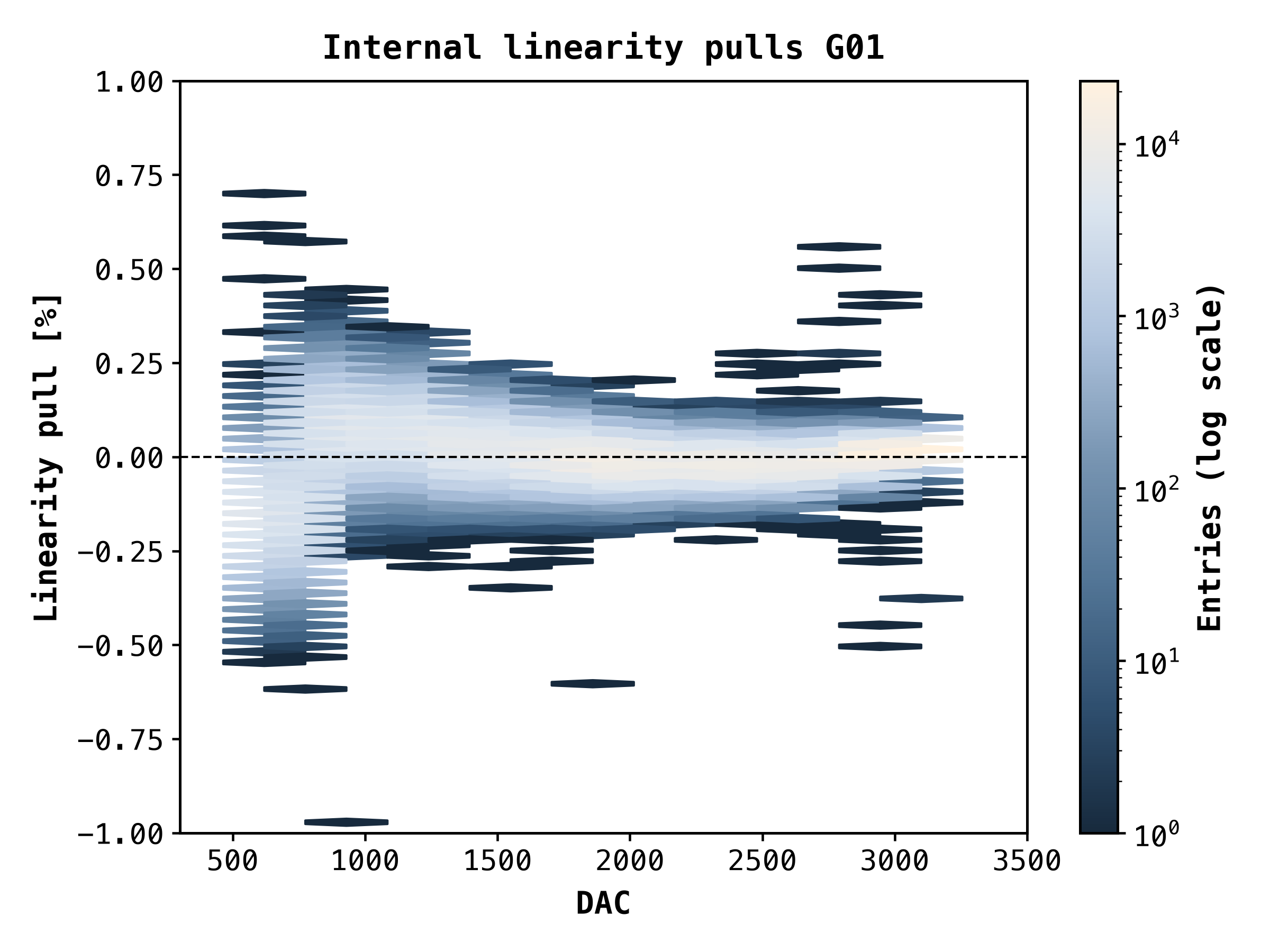}\quad
    \includegraphics[width=0.45\linewidth]{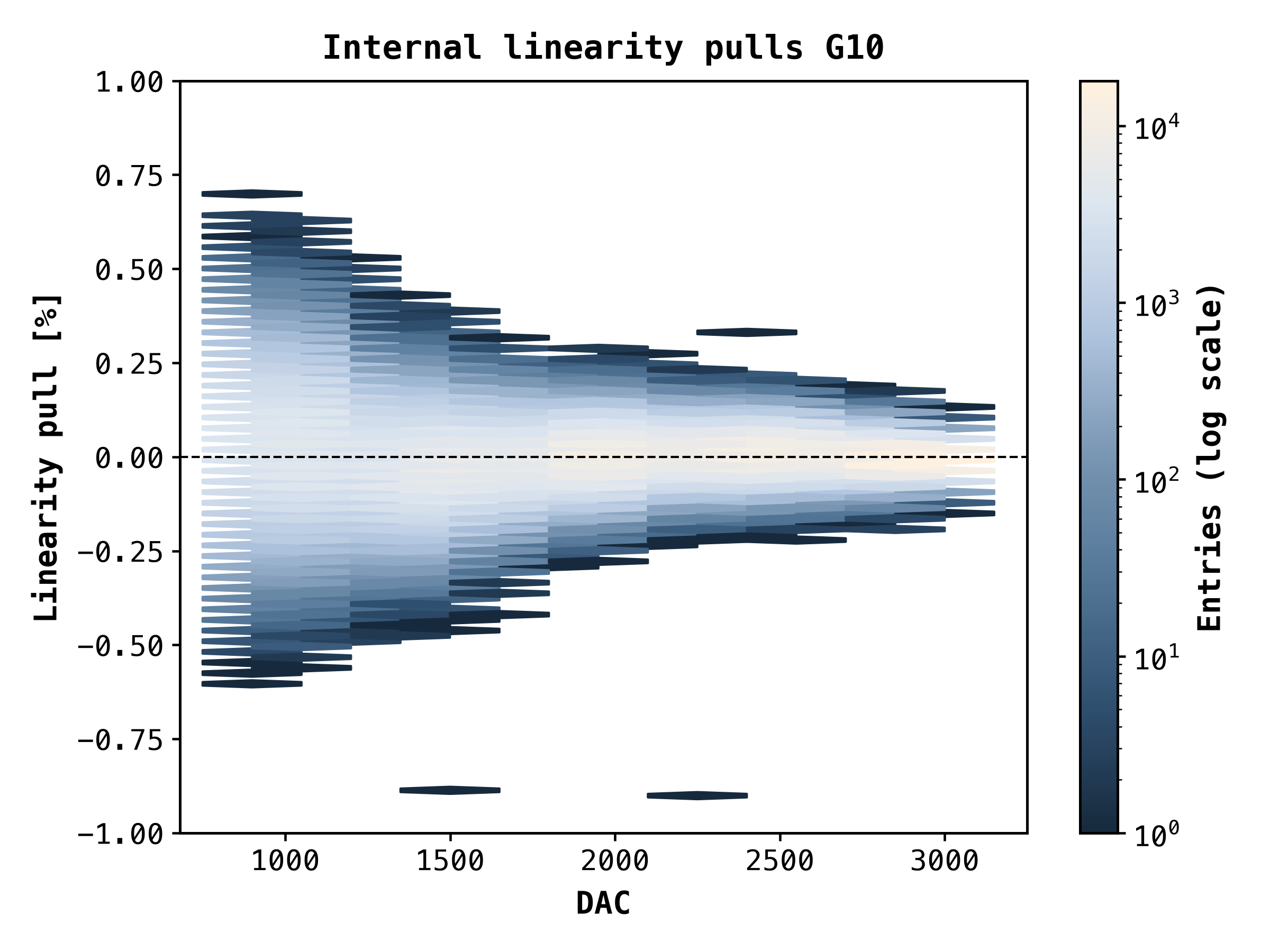}
    \vspace{-6pt}
    \caption{Internal-linearity pulls as a function of the pulse \acs{DAC} setting for gain~1 (left) and gain~10 (right).}   
    \label{fig:pulls}
\end{figure}

Internal linearity is checked with
ten amplitudes in high and low gain.
Figure~\ref{fig:pulls} shows the
pulls of the reconstructed amplitudes
with respect to the fitted response.
The pulls are concentrated near zero,
with amplitude-dependent tails that
are a diagnostic of departures from
linearity.
Seven cards in the analysed sample
failed this test.

\begin{figure}[htbp]
    \centering
    \includegraphics[width=0.95\linewidth]{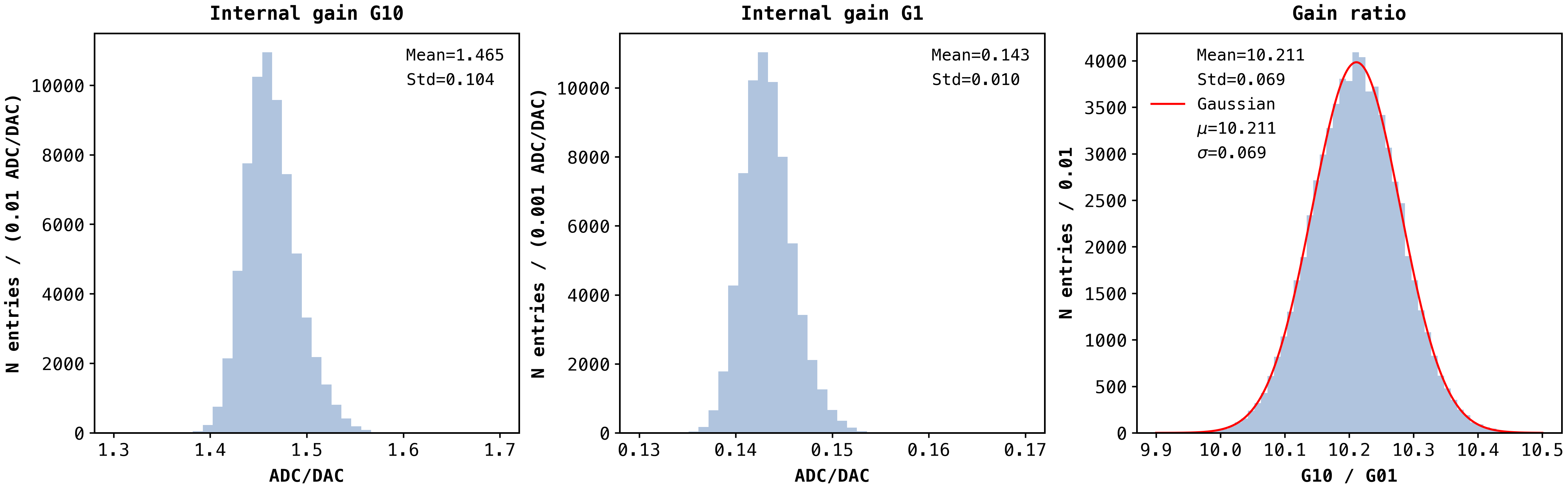}
    \vspace{-6pt}
    \caption{
    Distributions of the measured internal gains for gain~10 (left) and gain~1 (centre), and the corresponding gain ratios (right).}
    \label{fig:gains}
\end{figure}

Internal calibration uses ten
amplitudes of the \ac{CATIA}
test-pulse generator, covering the
full high-gain range and the
overlapping low-gain region, with
40 points per amplitude and gain.
Linear fits give the two channel
gains, used to calculate the
gain ratio.
The measurement is repeated for
$R_{\mathrm{TIA}}=\qtylist{340;400;500}{\ohm}$
with $C_1$ enabled.
For the \qty{340}{\ohm} resistor, which is 
foreseen for use at the CMS, Figure~\ref{fig:gains} 
shows compact gain distributions and a gain 
ratio of $\num{10.21 \pm 0.07}$.

\section{Discussion and conclusion}
\label{sec:conclusion}

The \ac{CERN} reception tests check the
full \ac{VFE} readout chain, from
electrical operation and clock
locking to baseline adjustment,
noise and pulse reconstruction.
A reconstructed fixed-amplitude
internal test pulse in the
expected range demonstrates full
functionality of the readout
channels.
The high production yield, close
to 99\%, together with the
uniformity of the measured
performance parameters, provides
a comfortable margin for the
complete refurbishment of the
\ac{EB}.

These results refer to \ac{CERN}
reception of the analysed sample
and do not include \ac{ESS} or
post-\ac{ESS} external
calibration.
After \ac{ESS} and final
calibration, the production
margin beyond the 12240 cards
required for the \ac{EB} will
allow quality-based selection
before installation.


\acknowledgments
\vspace{-6pt}
The authors thank Alexandre Dolgopolov
and Alexander Singovski from University
of Notre Dame, USA, for providing the
VICE++ cards used in the \ac{VFE} test
setup. Patricia Rebello Teles warmly thanks the support of CNPq via the grant 201440/2026-1. Marko Kovač gratefully acknowledges the support of the Croatian Science Foundation under project number IP-2025-02-6643. The authors would like to acknowledge the contributions of all colleagues involved in the VFE card testing.
This research was supported by the
Swiss National Science Foundation
(SNF) via the grant SNF FLARE 20FL20\_232672.

\end{document}